\pdfoutput=1
\newcommand*{\ATLASLATEXPATH}{}
\documentclass[cernpreprint,USenglish,texlive=2016,txfonts,texmf]{atlasdoc}

\usepackage[subfigure]{atlaspackage}

\usepackage{atlasbiblatex}
\usepackage{atlascontribute}
\usepackage{atlasphysics}

\graphicspath{{logos/}{figures/}}

\usepackage{vbspaper-defs}
\usepackage{dcolumn}
\newcolumntype{?}{D{,}{\,\pm\,}{-1}}
\newcolumntype{!}{D{,}{\,\pm\,}{4.4}}

\AtlasTitle{Search for anomalous electroweak production 
of $WW/WZ$ in association with a high-mass dijet system
in $pp$ collisions at $\rts=8~\tev$ with the ATLAS detector}

\author{The ATLAS Collaboration}

\PreprintIdNumber{CERN-EP-2016-171}

 \AtlasJournalRef{Phys.\ Rev.\ D 95 (2017) 032001}
\AtlasDOI{10.1103/PhysRevD.95.032001}

\AtlasAbstract{%
  A search is presented for anomalous quartic gauge boson couplings in vector-boson scattering.
The data for the analysis correspond to $\ourLumiNoError$ of $\rts=8~\tev$ $pp$ collisions,
and were collected in 2012 by the ATLAS experiment at the Large Hadron Collider.  
The search looks for the production of $WW$ or $WZ$ boson pairs accompanied
by a high-mass dijet system, with one $W$ decaying leptonically, and a $W$ or $Z$ 
decaying hadronically. The hadronically decaying $W/Z$ is reconstructed as 
either two small-radius jets or one large-radius jet using jet substructure techniques.  
Constraints on the anomalous quartic 
gauge boson coupling parameters $\alpha_4$ and $\alpha_5$ are set by fitting the 
transverse mass of the diboson system, and the resulting 95\% confidence intervals 
are $\aFourLoObs<\alpha_4<\aFourHiObs$ and $\aFiveLoObs<\alpha_5<\aFiveHiObs$.
}

\hypersetup{pdftitle={ATLAS document},pdfauthor={The ATLAS Collaboration}}

\begin{document}

\maketitle

\tableofcontents

%-------------------------------------------------------------------------------
\section{Introduction}
\label{sec:intro}
One of the main goals of the LHC experiments is to elucidate
the mechanism of electroweak 
symmetry breaking (EWSB).  In the Standard Model (SM), EWSB is explained by
the Brout--Englert--Higgs mechanism~\cite{Englert:1964et,Higgs:1964pj,Guralnik:1964eu}.   
Although many measurements have been made 
of the properties of the Higgs boson, more information is needed for a complete picture 
of EWSB.  Vector-boson scattering (VBS) is a key probe of EWSB, since 
it is sensitive to interactions between the longitudinal components of the 
gauge bosons.

ATLAS and CMS have recently presented results of 
VBS searches~\cite{STDM-2013-06,CMS-SMP-13-015,Aad:2016ett}, 
and although the searches in the $\ssWW$ channel are reaching sensitivity
to the Standard Model (SM) VBS process, an observation
has not yet been claimed.
 However, even without an observation of the SM process,
these analyses have been able to constrain 
physics beyond the SM (BSM).  

A common way of parameterizing BSM physics in VBS is through a 
low-energy effective theory~\cite{Longhitano:1980tm}.
Such an approach avoids having to choose a specific BSM theory, and
is particularly well suited if the energy scale
of the BSM physics is too high for the new resonances of the theory
to be observed directly.  In this kind of framework, VBS can be modified 
by anomalous quartic gauge couplings (aQGCs).  Searches for aQGCs have
been performed by the LEP experiments~\cite{delphi1, L3_1, L3_2, opal1, opal2, opal3}, 
 D0~\cite{d0aQGC}, and
the LHC experiments~\cite{STDM-2013-06,STDM-2013-05,Aad:2016ett, atlas4, ssWW8TeVPRD, WWW8TeVATLAS, CMS-SMP-13-009, CMS-SMP-13-015, cms4}.
 A typical prediction of aQGCs is an 
enhancement of the VBS cross-section at high transverse momentum ($\pT$)
of the vector bosons and at high invariant mass of the diboson system.

Experimentally, VBS is characterized by the presence of a pair of vector
bosons ($W$, $Z$, or $\gamma$) and two forward jets with a large
separation in rapidity and a large dijet invariant mass.
Previous searches for aQGCs in VBS have focused on channels involving
leptonic boson decays ($\Wlv$ and $\Zll$)~\footnote{Unless otherwise
noted, $\ell=e,\mu$ in this paper.} and photons.
The $\Vqq\Wlv$ channel ($V=W,Z$), however, offers some interesting
advantages.
The $\Vqq$ branching fractions are much larger than the leptonic
branching fractions.
Also, the kinematics of $\Vqq\Wlv$ are easier to reconstruct than $\Wlv\Wlv$ 
because there is one less neutrino in the final state, which enhances
the sensitivity to aQGC-dependent kinematic effects.
In addition, the use of jet substructure techniques allows good 
reconstruction efficiency in the high-$\pT$ region, which is the most sensitive
to aQGCs. 
The main challenge of the $\Vqq\Wlv$ channel is the 
presence of large backgrounds from $\Wjet$ and $\ttbar$ events.  These
backgrounds make a SM VBS measurement in this channel very
challenging because it is difficult to achieve a favorable
signal-to-background ratio.   On the other hand, an aQGC search is less sensitive 
to these backgrounds  because it is possible to find regions of phase space where the aQGC signal
is greatly enhanced over the SM processes, resulting in large signal-to-background ratios.
This motivates a search for aQGCs in the $\Vqq\Wlv$ channel. 

In this analysis, the approach used in Ref.~\cite{Alboteanu:2008my} is 
adopted, which parameterizes aQGCs by adding two new operators to the SM:
\begin{eqnarray}
\alpha_{4} {\cal L}_{4} & = & \alpha_{4}\, \text{tr}[\matr{V}_{\mu}\matr{V}_{\nu}]\, \text{tr}[\matr{V}^{\mu}\matr{V}^{\nu}]\,, \nonumber \\
\alpha_{5} {\cal L}_{5} & = & \alpha_{5}\, \text{tr}[\matr{V}_{\mu}\matr{V}^{\mu}]\, \text{tr}[\matr{V}_{\nu}\matr{V}^{\nu}]\,,
\end{eqnarray}
where the $\matr{V}_{\mu}$  field is related to the gauge boson fields. 
The SM (including the Higgs boson) is recovered when $\alpha_4=\alpha_5=0$. This model, with the simple addition
of two aQGC parameters to the SM, is not an ultraviolet-complete theory, and it
must be modified to prevent unitarity violation at high energies. In this analysis,
the K-matrix unitarization method~\cite{Alboteanu:2008my} is applied in order to ensure
that the aQGCs do not lead to the violation of unitarity.  This
aQGC parameterization and unitarization method was also adopted 
in Refs.~\cite{STDM-2013-06,Aad:2016ett}.
Both the $\alpha_4$ and $\alpha_5$ parameters lead to similar modifications of the 
VBS phenomenology: an increase in the cross-section and changes in the kinematics,
most notably an enhancement of VBS at high $VV$ invariant mass.

This paper presents a study of the production of $\Vqq\Wlv$ accompanied by a high-mass
dijet system, in a phase space optimized for sensitivity to aQGCs.  The $\Vqq$
system is reconstructed in two different ways: as two small-radius jets, or as a single
large-radius jet making use of jet substructure.  A search for
aQGC effects is performed using the transverse-mass distribution of the diboson system.

%-------------------------------------------------------------------------------

%-------------------------------------------------------------------------------
\section{ATLAS detector}
\label{sec:detector}
The ATLAS detector~\cite{PERF-2007-01} has a cylindrical 
geometry,\footnote{ATLAS uses a right-handed coordinate system with its origin at the nominal interaction point (IP) in the center of the detector and the $z$-axis along the beam pipe. The $x$-axis points from the IP to the center of the LHC ring, and the $y$-axis points upward. Cylindrical coordinates $(r,\phi)$ are used in the transverse plane, $\phi$ being the azimuthal angle around the $z$-axis. The pseudorapidity is defined in terms of the polar angle $\theta$ as $\eta=-\ln\tan(\theta/2)$.} and 
consists of several layers of subdetectors around the interaction point.
The innermost layer, the inner detector (ID) provides charged-particle
tracking for $|\eta|<2.5$.  The ID is surrounded by a superconducting solenoid providing a 2~T magnetic field, and the solenoid in turn is surrounded by a
liquid-argon (LAr) electromagnetic (EM) calorimeter that provides coverage
in the range $|\eta|<3.2$.  A scintillator-tile calorimeter provides
hadronic measurements for $|\eta|<1.7$ and LAr calorimeters in the 
forward region provide additional EM and hadronic measurements up to $|\eta|=4.9$.
A muon spectrometer (MS) surrounds the calorimeters and makes use of a toroidal
magnetic field.  The MS provides tracking capabilities for $|\eta|<2.7$ and triggering for
$|\eta|<2.4$.  Events are selected for offline processing using a three-level trigger
system.

%-------------------------------------------------------------------------------

%-------------------------------------------------------------------------------
\section{Data and Monte Carlo samples}
\label{sec:samples}
This analysis uses $\ourLumi$~\cite{Aaboud:2016hhf} of  $8~\tev$ $pp$ collision data recorded by 
the ATLAS detector in 2012.
Events used in this analysis are required to pass one of several single-lepton triggers.  
One set of triggers requires an isolated electron or muon with $\pT>24~\gev$.
Another set of triggers requires an electron (muon) with  $\pT>60\, (36) ~\gev$, without the isolation requirement.

\begin{figure}[htbp]
\begin{center}
 \includegraphics[width=0.4\textwidth,keepaspectratio]{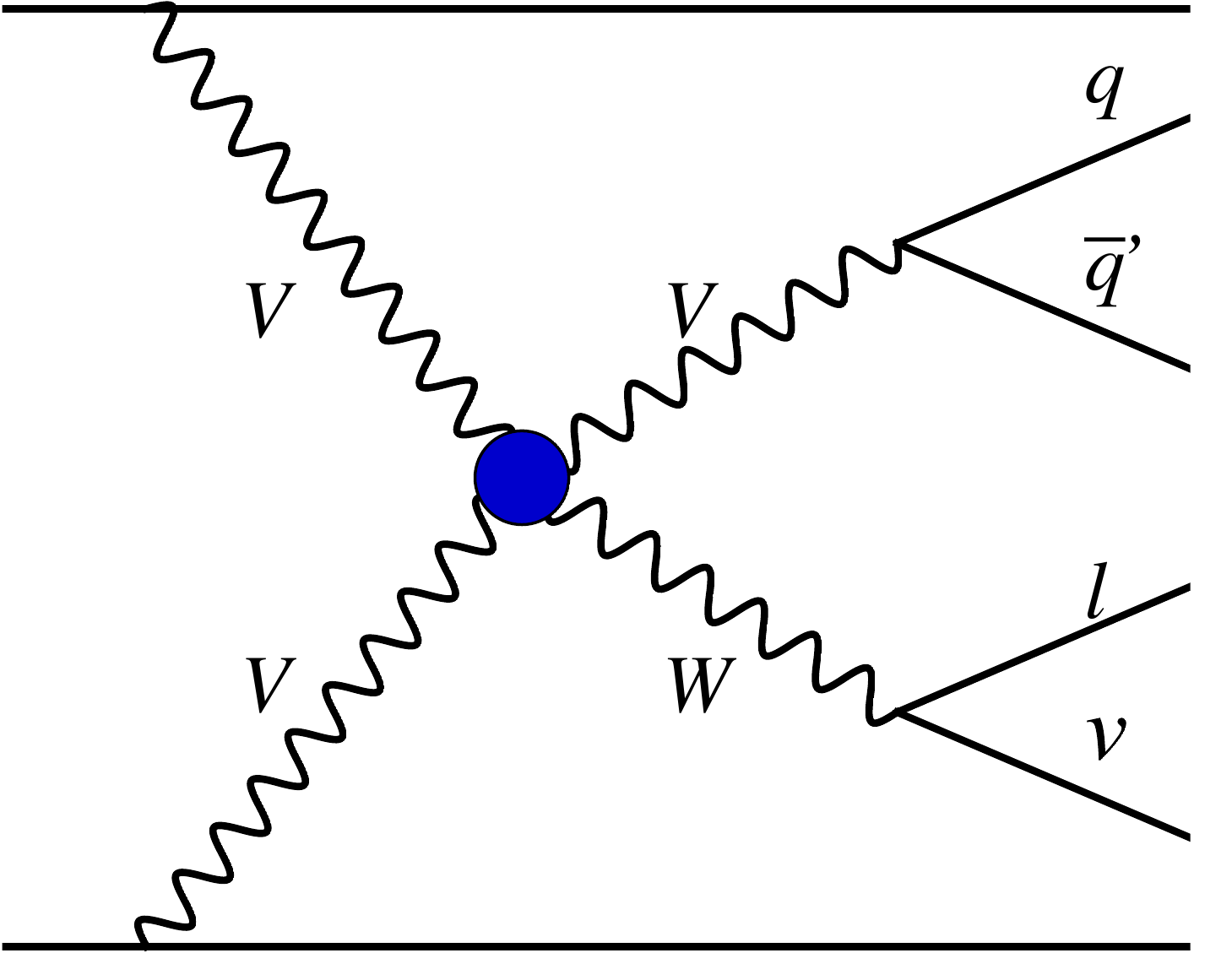}
\caption{
{A VBS diagram that contributes to \signal\ production.  This analysis searches for modifications
of the quartic gauge couplings.}
}
\label{fig:feynman}
\end{center}
\end{figure}

This analysis searches for anomalous contributions to electroweak (EWK) production
of two vector bosons plus two jets, which is hereafter referred to as ``\signal.''
The \signal\ process is modeled with Monte Carlo (MC) samples that include $\Vqq\ell\nu+2~\textrm{parton}$ and 
$\Vqq\ell^+\ell^-+2~\textrm{parton}$ production, and include all the
purely electroweak (i.e., $\mathcal{O}(\alpha_{\text{EWK}}^6)$) tree-level diagrams that contribute
to these final states. The \signal\ process definition includes both the VBS and non-VBS diagrams 
because the VBS-only process cannot be defined in a 
gauge-invariant way~\cite{Accomando:2006mc}.  One example of the 
\signal\ diagrams is shown in Figure~\ref{fig:feynman}.
Production of $\Vqq\ell\nu+2~\textrm{parton}$ and $\Vqq\ell^+\ell^-+2~\textrm{parton}$ can
also occur through diagrams that are $\mathcal{O}(\alpha_{\text{EWK}}^4\alphas^2)$ at tree level,
but such processes are not affected by quartic gauge couplings, and are not considered as \signal,
but rather are included in the diboson background described below.
In the \signal\ MC sample definition, ``$\ell$'' includes tau leptons, in order to account for
contributions from $\tau\to(e/\mu)+X$ decays that could pass the event selection.

The \signal\ process is modeled with \whizard~v2.1.1~\cite{Kilian:2007gr,Moretti:2001zz}, 
complemented by the \PYTHIA8~\cite{Sjostrand:2007gs} parton shower, fragmentation, and hadronization modeling, 
and using the CT10 parton distribution function (PDF) set~\cite{Lai:2010vv}.
\whizard is used to generate both the SM samples and samples with non-zero aQGC values.
The samples use dynamic factorization and renormalization scales equal to the diboson invariant mass.  
The SM and aQGC samples are normalized using  the leading-order (LO) cross-sections from \whizard.

The $\Wjet$ and $\Zjet$ backgrounds are modeled using
\SHERPA~v1.4.1~\cite{Gleisberg:2008ta,Hoeche:2009rj,Gleisberg:2008fv,Schumann:2007mg},  
with up to four partons in the matrix element. The CT10 PDF set is used.    
These samples are normalized using next-to-next-to leading-order (NNLO)
inclusive cross-sections obtained from FEWZ~\cite{Anastasiou:2003ds}.
These samples do not contain electroweak production of $\Wjet$ (for example, $W$-production through vector-boson fusion),
which is modeled separately with \SHERPA~v1.4.3 and the CT10 PDF set.

Backgrounds from $\ttbar$ events and   single-top-quark events in the $Wt$- and $s$-channels
are generated with  \POWHEGBOXmine~\cite{Nason:2004rx,Frixione:2007vw,Alioli:2010xd,Alioli:2009je,Re:2010bp}
using the CT10 PDF set.
Parton showering is done with \PYTHIA v6.426~\cite{Sjostrand:2006za} 
using the P2011C set of tuned parameters (P2011C tune)~\cite{Skands:2010ak}.  
The $t$-channel single-top-quark process is modeled with AcerMC~\cite{Kersevan:2004yg} plus \PYTHIA v6.426 with the 
P2011C tune and the CTEQ6L1 PDF set~\cite{Pumplin:2002vw}. 
The $\ttbar$ samples are normalized using the NNLO cross-section 
 including resummation of next-to-next-to-leading logarithmic (NNLL) soft gluon terms, calculated
with top++2.0~\cite{Cacciari:2011hy,Beneke:2011mq,Baernreuther:2012ws,Czakon:2012zr,Czakon:2012pz,Czakon:2013goa,Czakon:2011xx}.
The single-top-quark samples are normalized using NLO+NNLL calculations~\cite{Kidonakis:2011wy,Kidonakis:2010tc,Kidonakis:2010ux}.

Backgrounds from diboson ($WW$, $WZ$, and $ZZ$) production  are modeled with 
 \SHERPA~v1.4.3 using the CT10 PDF set. 
These samples are normalized using NLO cross-sections~\cite{Campbell:2010ff}.
These background samples do not overlap with the \signal\ samples, since the former do not include 
purely electroweak production of dibosons in association with two jets.

The $W\gamma$ background is modeled with 
\ALPGEN~\cite{Mangano:2002ea} interfaced with 
\HERWIG~v6.520.2~\cite{Corcella:2000bw} and \JIMMY~\cite{Butterworth:1996zw},
using the CTEQ6L1 PDF set and AUET2 tune~\cite{ATL-PHYS-PUB-2011-008}.
The $Z\gamma$ background is modeled with \SHERPA~v1.4.1 and the CT10 PDF set.

The MC samples are passed through the ATLAS detector simulation~\cite{SOFT-2010-01}, which 
is based on GEANT4~\cite{GEANT4}. Some of the samples are passed through a fast simulation
that uses a parameterization of the electromagnetic and hadronic calorimeters. 
The simulated hard-scattering processes are overlaid 
with minimum-bias events, in order to model additional $pp$ interactions in the events (pile-up).  
The simulated events are reweighted in order to better match the number of interactions per bunch 
crossing observed in data.

%-------------------------------------------------------------------------------

%-------------------------------------------------------------------------------
\section{Object selection}
\label{sec:objselection}
The analysis selects events with exactly one lepton (either an electron or muon), missing transverse
momentum, and either four small-radius jets or two small-radius jets and one large-radius jet.

``Loose'' electron candidates are reconstructed by matching energy deposits in the EM calorimeter to
tracks in the ID.  They must have transverse energy $\ET>15~\gev$ and $|\eta|<2.47$, excluding
the transition region between the barrel and endcap calorimeters $1.37<|\eta|<1.52$. Their longitudinal impact parameter
with respect to the primary vertex, $z_0$, must satisfy $|z_0 \sin{\theta}|<0.5$~mm, and their transverse impact parameter $d_0$ must
satisfy $\dzsig<5$, where $\sigmadz$ is the uncertainty in $d_0$. This reduces electron candidates from heavy-flavor decays.
  Also, they must satisfy
``medium'' cut-based identificaton criteria from Ref.~\cite{ATLAS-CONF-2014-032} that are based on the calorimeter shower
shape and track variables, and which are designed to reduce fake electron candidates from backgrounds such as jets.
The candidates are rejected if they are within $\DR=\sqrt{(\Delta\eta)^2 + (\Delta\phi)^2}=0.1$ of a ``good'' muon, defined below.

``Loose'' muon candidates are found by combining tracks from the ID with tracks from the MS.  They must 
have a transverse momentum $\pT>15~\gev$,  $|\eta|<2.4$, and $|z_0\sin{\theta}|<0.5$~mm. They are
also required to have a certain number of hits in each layer of the ID.

``Good'' lepton candidates are a subset of loose lepton candidates that satisfy additional criteria.  
Good electrons must satisfy the ``tight'' cut-based identification criteria from Ref.~\cite{ATLAS-CONF-2014-032}.
Good muons must have $\dzsig<3$. Electrons and muons must both pass isolation requirements, in order to reduce contributions
from jets misreconstructed as electrons, or from leptons originating from heavy-flavor hadronic decays.
Electrons (muons) must have $\caloiso<0.14~(0.07)$ and $\trackiso<0.07~(0.07)$.  Here $\caloiso$
is the scalar sum of the $\ET$ of energy deposits in the calorimeter within a cone of size $\DR=0.3$ around the lepton candidate
(excluding the lepton candidate itself), divided by the electron $\ET$ or muon $\pT$. The quantity $\trackiso$
is calculated as the scalar sum of the $\pT$ of the tracks within $\DR=0.3$ of the lepton candidate (but excluding the
lepton candidate), divided by the electron $\ET$ or muon $\pT$.

Small-radius jets (hereafter ``small-$R$'' jets) are reconstructed using the anti-$k_t$ 
algorithm~\cite{Cacciari:2008gp} with radius parameter $0.4$.  Small-$R$ jets must have $\pT>30~\gev$
and $|\eta|<4.5$, and must be separated from lepton candidates by at least $\DR=0.3$.  
Small-$R$ jets with $\pT<50~\gev$ and $|\eta|<2.4$ must also have 
a ``jet vertex fraction''~\cite{ATLAS-CONF-2013-083} 
with absolute value greater than 0.5, in order to reject jets from other simultaneous $pp$ 
collisions.

Large-radius  (``large-$R$'') jets are reconstructed using the 
Cambridge--Aachen algorithm~\cite{Dokshitzer:1997in}
with radius $1.2$, and  are ``groomed'' using a mass-drop filtering 
algorithm~\cite{Butterworth:2008iy} with filtering criteria $\mu_\textrm{frac}<0.67$ and $y_f>0.09$.
This algorithm selects jets that contain substructure consistent with a two-body decay.
Large-$R$ jets must have $\pT>200~\gev$ and $|\eta|<1.2$, and be separated from lepton candidates
by at least $\DR=1.2$.

The missing transverse momentum $\vecmet$ is calculated as the negative vector sum of the
$\pT$ of all the objects in the events.  The $\pT$ of electrons, muons, photons, and jets
are taken from reconstructed objects, and a ``soft term'' accounting 
for the transverse energy of calorimeter clusters not associated
with any reconstructed object is also included~\cite{PERF-2011-07}.

\section{Event selection}
\label{sec:eventselection}

In order to ensure that selected events are due to proton--proton collisions, each event is required
to have at least one reconstructed vertex with at least three tracks having $\pT>400~\mev$.
Events must have exactly one ``good'' electron or muon with $\pT(\ell)>30~\gev$, and events containing 
any additional ``loose'' electrons or muons are vetoed.  The $\met$ in the event must be greater than $30~\gev$. 
The leptonically decaying $W$ candidate, $\Wlep$, is formed by the four-momentum sum of
the lepton and the missing momentum, where the $z$-component of the missing momentum is inferred by requiring
the invariant mass of $\Wlep$ to be equal to the nominal $W$ mass of 
$80.4~\gev$~\cite{Agashe:2014kda}.  

For reconstructing the hadronic portion of the event, two different selection criteria are used.  A
``\resolved'' selection is developed that reconstructs the hadronically decaying $W/Z$ candidate ($\Vhad$)
as two small-$R$ jets ($\Vtojj$), whereas a ``\merged'' selection reconstructs the $\Vhad$ as a single
large-$R$ jet ($\VtoJ$).  

For the \resolved\ selection, the event must have at least four small-$R$ jets.
The $\Vhad$ candidate is formed from the two jets that have $\mjj$ closest to the 
nominal $W$ mass, unless there are multiple jet pairs with $\mjj$ within $15~\gev$  of the $W$ mass, 
in which case $\Vhad$ is chosen from among these jet pairs, using an algorithm that favors
jet pairs with two high-$\pT$ jets.
From the remaining small-$R$ jets, the two that have the highest $\mjj$ are chosen as the
``tagging'' jets.

For the \merged\ selection, the event must have at least one large-$R$ jet, which represents
the $\Vhad$ candidate.  In the case of multiple large-$R$ jets, the one with mass closest to the
nominal $W$ mass is taken as the $\Vhad$ candidate.  The event must also have at least two 
small-$R$ jets that each have $\DR(\jet,\Vhad)> 1.2$.  Among these small-$R$ jets, the two
with the highest $\mjj$ are chosen as the tagging jets.

In both the \resolved\ and \merged\ selections, the $\Vhad$ candidate must have
$\cutmjjloR<m(\Vhad)<\cutmjjhiR~\gev$, and the invariant mass of the tagging jets must be
$\mjjTag>500~\gev$. The requirement on $m(\Vhad)$ favors the $WW$ component of the
\signal\ process over the $WZ$ component; however, the latter is only expected to contribute
$10-15\%$ of the total \signal\ events in the phasespace of this analysis, both for the SM
and for aQGC contributions.

In order to reduce the amount of background from \ttbar and single-top-quark processes, a restriction
is placed on the number of $b$-tagged jets in the event.  Small-$R$ jets are tagged as $b$-jets
using the ``MV1'' algorithm~\cite{ATLAS-CONF-2014-046,Aad:2015ydr} with a $b$-tag efficiency of 85\%.  In the \resolved\ selection,
the event is vetoed if (a) both of the jets associated with the $\Vhad$ candidate are $b$-tagged, 
or (b) if any other jet in the event is $b$-tagged.  The reason for not vetoing events that have only a
single $b$-tagged $\Vhad$-jet is to prevent \signal\ events with a $W\to cs$ decay from being vetoed
due to a  mistagged $c$-jet. 
  In the \merged\ selection, the event is
vetoed if any small-$R$ jet with $\DR(\jet,\Vhad)> 0.4$ is $b$-tagged.

The aforementioned event selection is designed to give a phase space with characteristics 
typical of VBS events,
and is referred to as the ``\looseVBS'' selection stage.
On top of the \looseVBS\ selection, additional selection criteria are applied that 
increase the sensitivity to aQGCs.  The minimum $\mjjTag$ value is increased to
 $\cutmjjtagR~\gev$ in both the \resolved\ and \merged\ selections.
In addition, events are required to have $\bosCen>\cutcentralR$, 
where $\bosCen$ is the \emph{boson centrality}, defined as
\begin{equation}
\bosCen = \min\{ \Delta\eta_-, \Delta\eta_+ \} \,,
\end{equation}
where $\Delta\eta_-  =   \min\{ \eta(\Vhad), \eta(\Wlep) \} - \min\{\eta_{\jettA}, \eta_{\jettB} \}$  and
$\Delta\eta_+  =   \max\{\eta_{\jettA}, \eta_{\jettB} \} -   \max\{ \eta(\Vhad), \eta(\Wlep) \}$. In these
equations, $\jettA$ and $\jettB$ refer to the two tagging jets.
The variable $\bosCen$ has large values when the tagging jets have large separation in $\eta$, and when the
two boson candidates are between the tagging jets in $\eta$.  The requirement
 $\bosCen>\cutcentralR$ implicitly forces $|\Delta\eta(\jettA,\jettB)|$ to be greater than $1.8$.
Furthermore, the $\pT$ of the $\Wlep$ candidate is required to be greater than 150~\gev.

For the \merged\ selection, the $\pT$-balance $\VVpTbal$ must be 
less than $\cutvvptbalF$, where
\begin{equation}
\VVpTbal = \frac{ | \vec{\pT}(\Vhad) + \vec{\pT}(\Wlep)| }{  \pT(\Vhad) + \pT(\Wlep) }\,.
\end{equation}
This requirement is based on the fact that the aQGC events are expected to have two bosons
produced roughly back-to-back.
For the \resolved\ selection, it is required that $\cosThWjj<\cutcosthR$, where 
$\thWjj$ is defined as the angle between the $\Vhad$ direction and one of the jets from the $\Vhad$ 
candidate. In this calculation, the $\Vhad$-jet direction is 
measured in the rest frame of the $\Vhad$, the $\Vhad$ direction
is measured in the $WV$ rest frame, and the $\Vhad$-jet used in this calculation is chosen to be whichever jet
gives $\cosThWjj>0$.  This $\cosThWjj$ requirement further improves aQGC sensitivity because 
aQGCs enhance the longitudinal 
polarization of the vector bosons at high $\pT$.  The thresholds for $\mjjTag$, $\bosCen$, $\VVpTbal$,
and $\cosThWjj$ were optimized for the best expected sensitivity to aQGCs.

To remove overlap between the \resolved\ and \merged\ selections, events that pass both selections are put in 
the \resolved\ category.
The search for aQGCs is performed by using the transverse mass of the diboson system, 
defined as
\begin{equation}
\mTWV = \sqrt{ (\ET(\Vhad) + \ET(\Wlep))^2 - ( p_x(\Vhad)+p_x(\Wlep) )^2 - ( p_y(\Vhad)+p_y(\Wlep) )^2}\,,
\end{equation}
where $\ET(\Vhad) =  E(\Vhad) \cdot \frac{\pT(\Vhad)}{p(\Vhad)}$ and 
$\ET(\Wlep) \equiv \ET(\ell) + \met$.  
The \merged\ category probes higher values of $\mTWV$ than the \resolved\ category. The signal
efficiency of the \resolved\ selection drops off rapidly over the range $600<\mTWV<800~\gev$,
and the \merged\ selection efficiency surpasses the \resolved\ selection efficiency for 
$\mTWV\gtrsim 700~\gev$.

Events are split up into three categories: $e^+$ and $\mu^+$ (\resolved\ selection),
$e^-$ and $\mu^-$ (\resolved\ selection), and the \merged\ selection.  The \resolved\ category is
split up by charge because the $\Wjet$ background and the aQGC signal are charge-asymmetric.  
The \merged\ category is not split up by lepton charge, because of the 
small expected event yield in this category.

%-------------------------------------------------------------------------------

%-------------------------------------------------------------------------------
\section{Background estimation}
\label{sec:backgrounds}
The main backgrounds in this analysis are due to $\Wjet$ and $\ttbar$ processes, 
with additional backgrounds from single-top-quark, non-electroweak diboson, 
$\Zjet$, and multijet events. All background predictions are taken from MC
simulation, except for the multijet background, which uses a data-driven 
prediction, and the $\Wjet$ background, which uses a MC prediction to 
which a data-driven scale factor is applied, as explained below.

\begin{figure}[htbp]
\begin{center}
\subfigure[]{
\label{subfig:WjetCR_resolve}
 \includegraphics[width=0.47\textwidth,keepaspectratio]{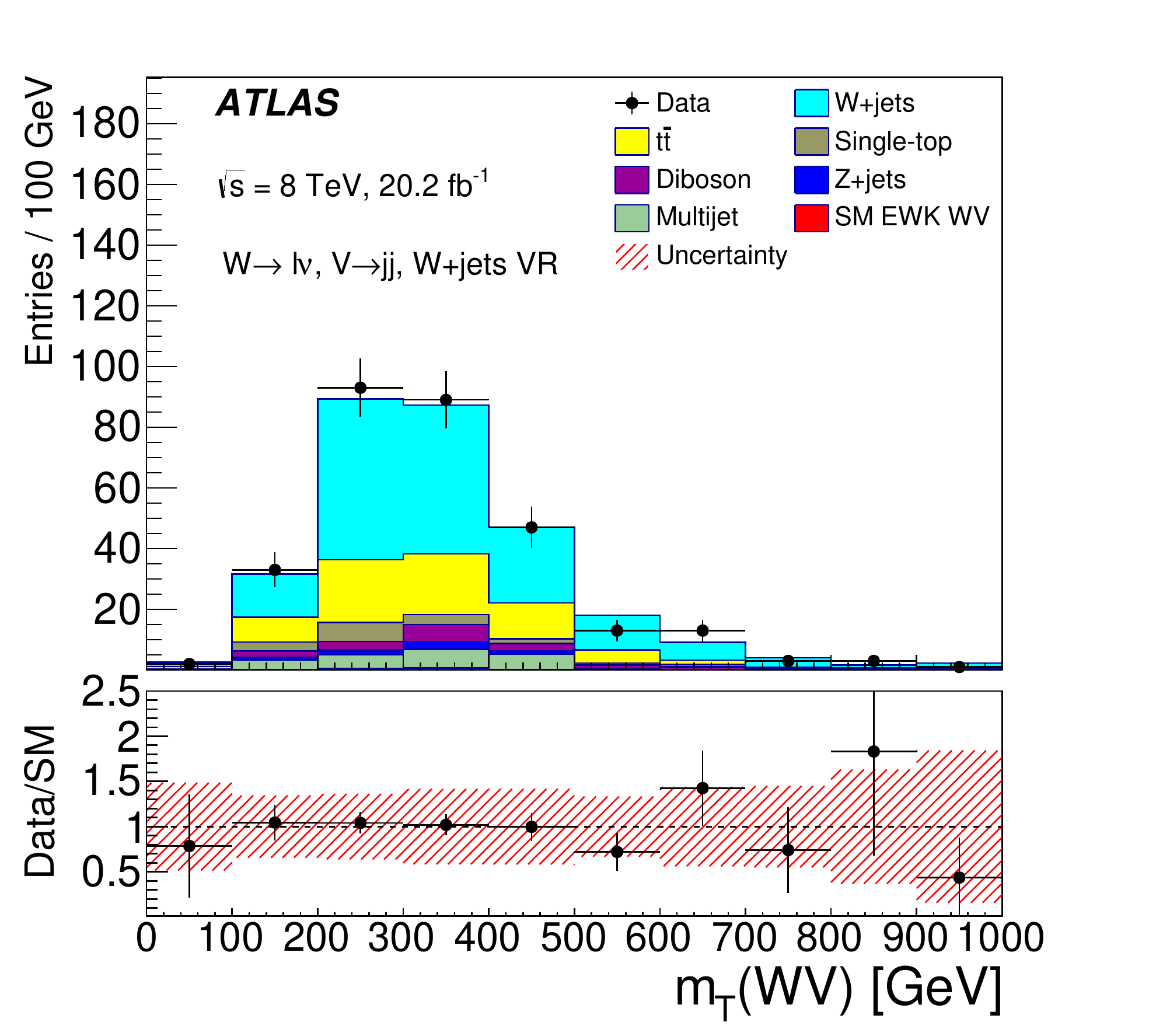}
}
\subfigure[]{
\label{subfig:WjetCR_fat}
 \includegraphics[width=0.47\textwidth,keepaspectratio]{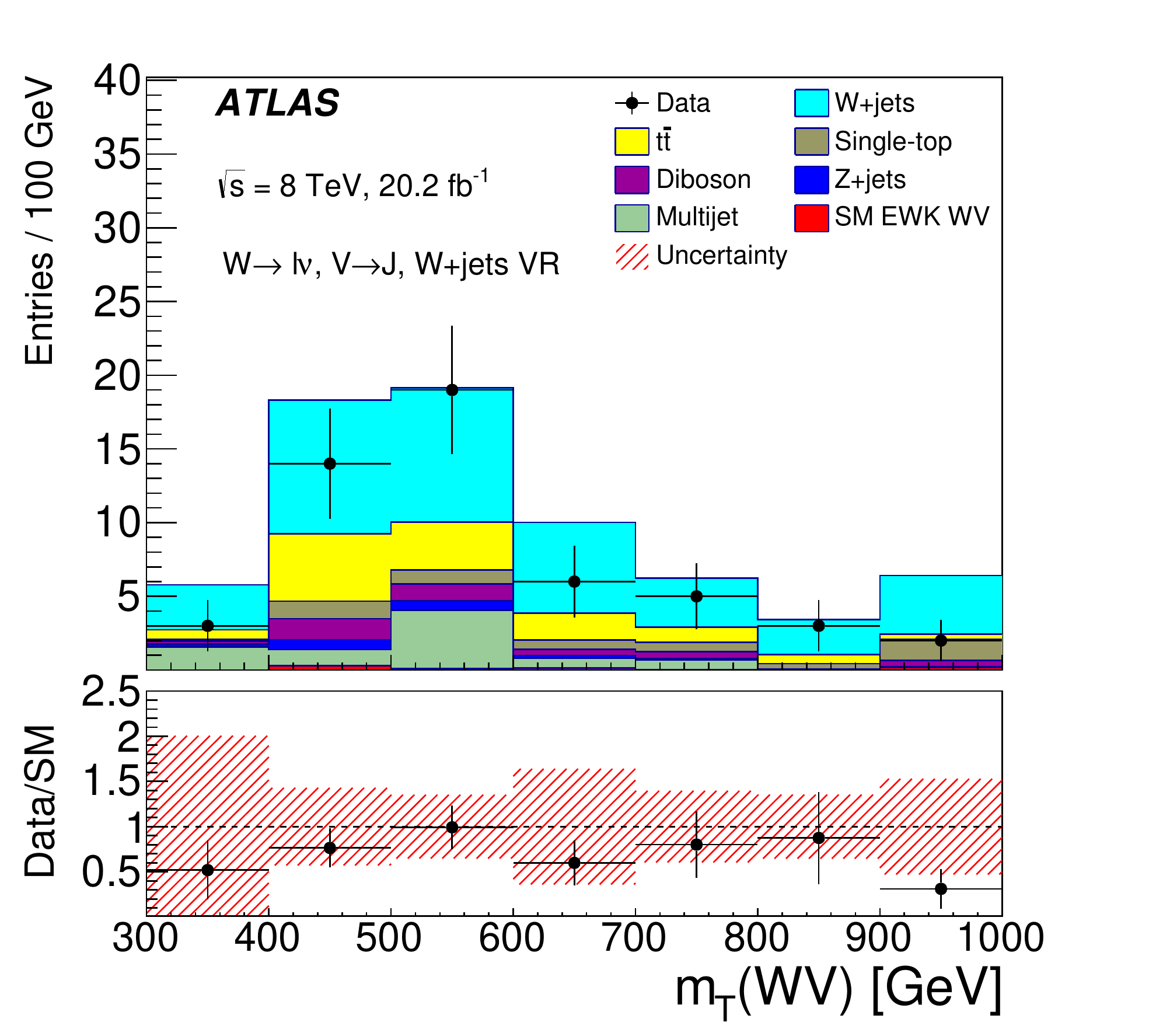}
}
\subfigure[]{
\label{subfig:TopCR_resolve}
 \includegraphics[width=0.47\textwidth,keepaspectratio]{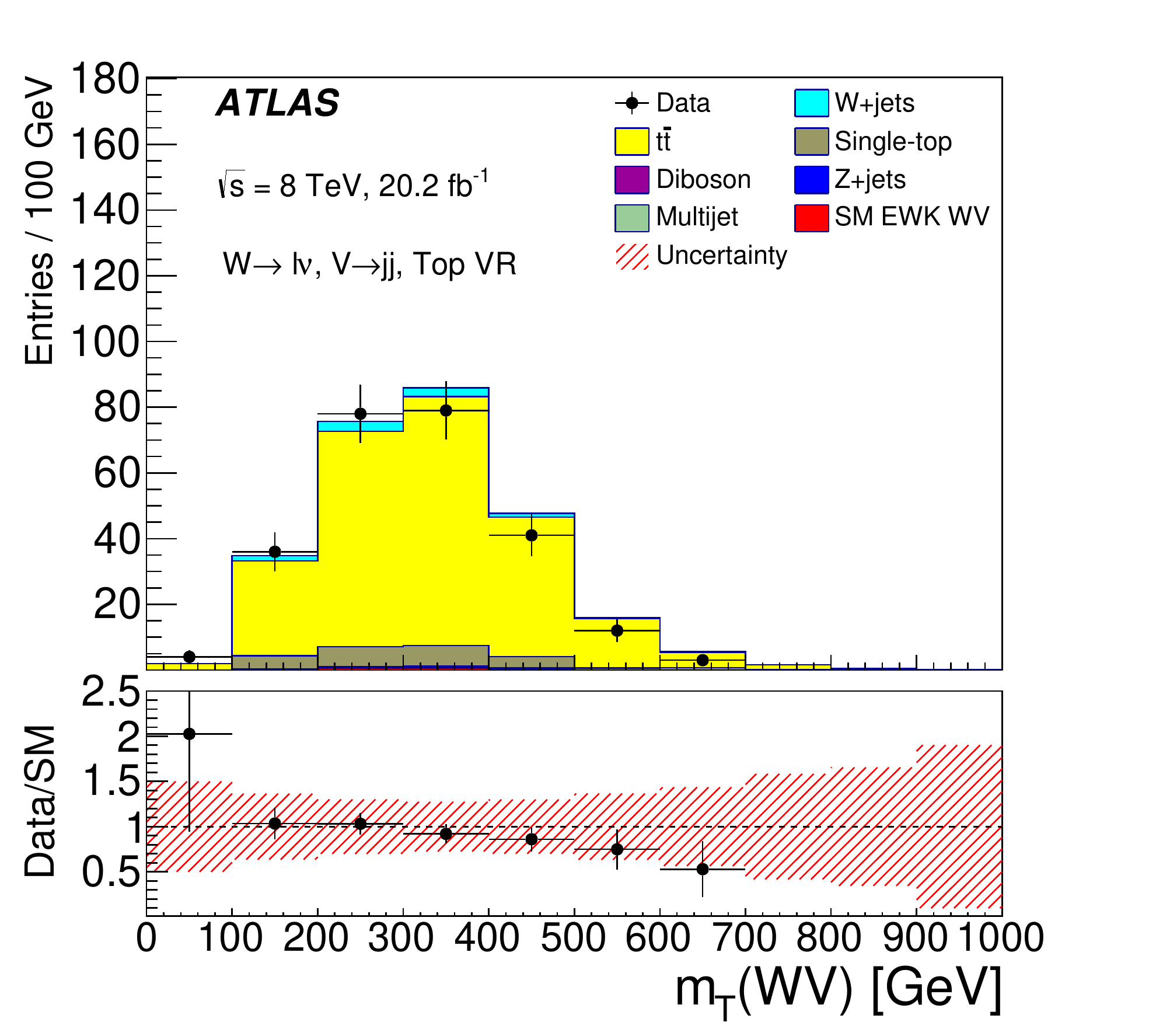}
}
\subfigure[]{
\label{subfig:TopCR_fat}
 \includegraphics[width=0.47\textwidth,keepaspectratio]{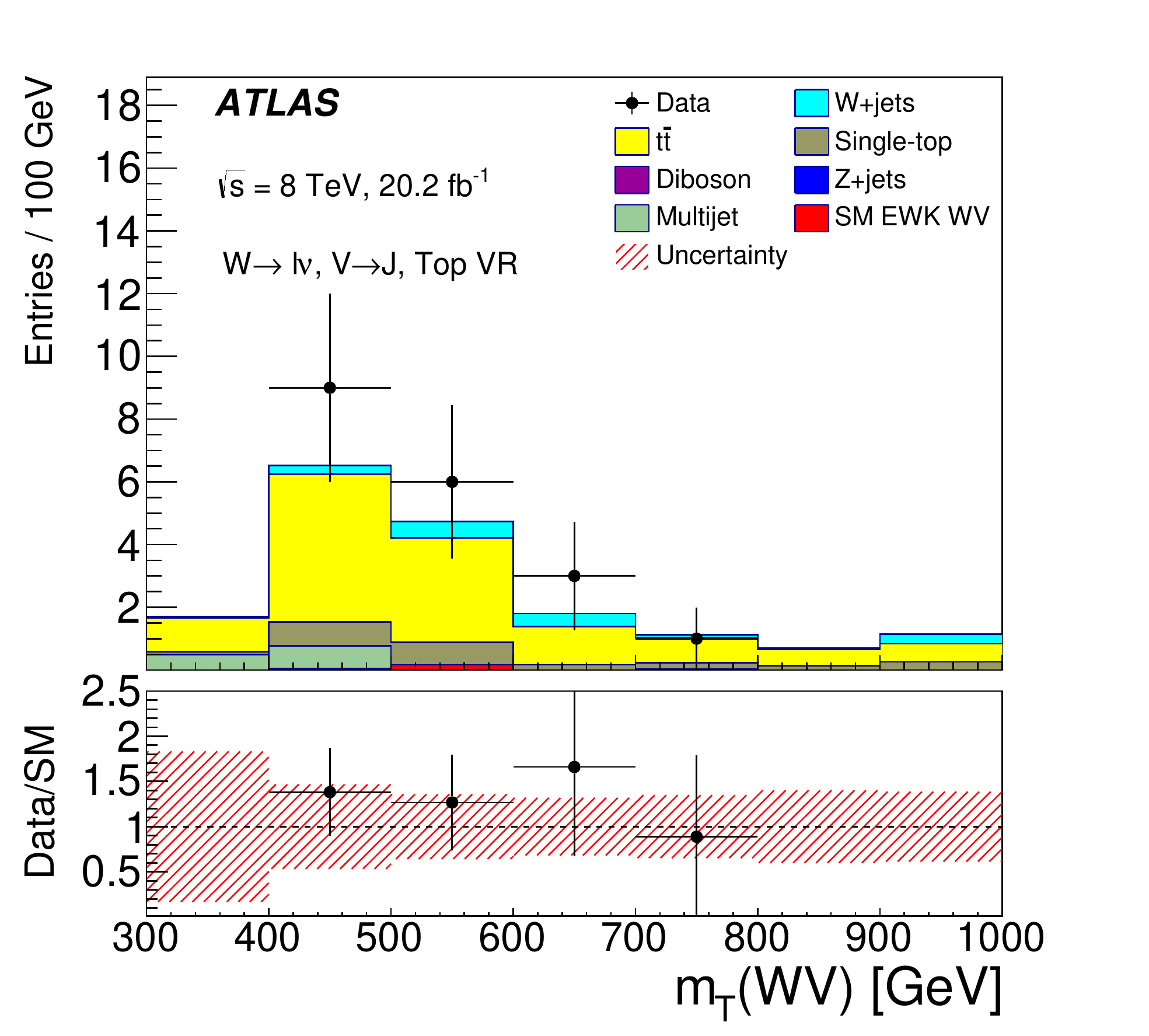}
}
\caption{
The top row 
shows the observed $\mTWV$ distribution in the $\Wjet$ validation region (VR), overlaid with the background prediction, for 
(a) the \resolved\ ($\Vtojj$) region, $e^+$, $e^-$, $\mu^+$, and $\mu^-$ combined;
and 
(b) the \merged\ ($\VtoJ$) region, $e^+$, $e^-$, $\mu^+$, and $\mu^-$ combined.
The bottom row shows the 
observed $\mTWV$ distribution in the Top-production VR,
again overlaid with the background prediction, for
(c) the \resolved\ region, $e^+$, $e^-$, $\mu^+$, and $\mu^-$ combined;
and
(d) the \merged\ region, $e^+$, $e^-$, $\mu^+$, and $\mu^-$ combined.
The last bin includes overflow.
}
\label{fig:WjetsAndTopCR}
\end{center}
\end{figure}

About half of the background events in this analysis are from 
$\Wjet$ production. Its modeling is checked using a control region 
(``loose $\Wjet$ CR'') defined using the ``loose VBS'' selection 
criteria, except that the $m(\Vhad)$ selection is 
inverted: $36<m(\Vhad)<\cutmjjloR~\gev$ or $m(\Vhad)>\cutmjjhiR~\gev$ for
the \resolved\ selection, and  $40<m(\Vhad)<\cutmjjloR~\gev$ or $m(\Vhad)>\cutmjjhiR~\gev$
for the \merged\ selection.
The background prediction is larger than the data in this region, which is attributed 
to an overestimate of the $\Wjet$ background by the MC simulation.  An average scale factor of
$0.82$ is derived for $\Wjet$ from this region, after subtracting 
the predictions for non-$\Wjet$ events. This constant scale factor is applied to the $\Wjet$ 
prediction in all three event categories.
 The $\Wjet$ modeling is 
cross-checked in a validation region (``$\Wjet$ VR'') defined using
the same selection as the signal region, except inverting the 
$m(\Vhad)$ selection.  The modeling of $\mTWV$ 
in this validation region is shown in Figures~\ref{subfig:WjetCR_resolve} and \ref{subfig:WjetCR_fat}.
The largest systematic uncertainties in the $\Wjet$ VR are jet uncertainties and
uncertainties in the modeling of the $\Wjet$ process, which are described
in Section~\ref{sec:systematics}.

Top-pair and single-top-quark production are the other major backgrounds
in this analysis.  Their modeling is checked in a validation region
(``Top VR'') that uses the same selection as the signal region,
except that the requirements on the number of $b$-tagged jets are inverted.
 The definition of a $b$-tagged jet is tightened for the Top VR; the MV1 algorithm is
used with a $b$-tag efficiency of 60\%. 
The data--MC comparison in the Top VR is shown in Figures~\ref{subfig:TopCR_resolve} and \ref{subfig:TopCR_fat}.
The largest systematic uncertainties in the Top VR are jet uncertainties and
uncertainties in the modeling of the $\ttbar$ process.
In both the $\Wjet$ VR and Top VR, the predicted event yields and $\mTWV$ distribution shapes
are consistent with those observed in data, within the systematic uncertainties.

Multijet processes are a fairly small background in this analysis. 
They can pass the event selection if a lepton from the decay
of a heavy-flavor hadron passes the lepton selection.  In the electron
channel, multijet events can also contribute due to jets misreconstructed 
as electrons.  They are modeled using a data-driven estimate as 
described below.
 
First, control regions are defined by event selections similar to those for the 
signal regions, but with modified lepton identification criteria, in order
to enrich the control regions in multijet backgrounds.  Leptons that satisfy
the modified identification criteria are referred to as ``bad'' leptons.   
For the muon channel, the impact-parameter criterion is inverted:
$\dzsig>3$.  For the electron channel, the
electron candidate must fail the ``tight'' cut-based identification but satisfy
the ``medium'' cut-based identification criteria from Ref.~\cite{ATLAS-CONF-2014-032}.
In addition, for both the electron and muon channels, the isolation criteria are 
modified:  $\caloiso>0.04$ and $\trackiso<0.5$.  
The shapes of the kinematic distributions ($\mTWV$, $\pTWlep$, $\met$) 
of the multijet background 
are obtained from the data in these control regions, after 
subtracting the MC predictions 
for the non-multijet backgrounds.

The multijet event yield is estimated by first performing a fit to the
$\met$ distribution of the data that pass the final event selection, but with 
the $\met>30~\gev$ and $\pTWlep>150~\gev$ criteria  removed.  
The final multijet yield estimate is then obtained by scaling this fit result
by the efficiency for multijet events to pass the $\met>30~\gev$ and 
$\pTWlep>150~\gev$ requirements. That efficiency is also estimated from a 
bad-lepton control region. The multijet estimate was cross-checked with
an alternative method that first applies the $\pTWlep>150~\gev$ selection, and
then obtains the multijet yield from a fit to the $\met$ distribution.

Remaining backgrounds originate from $\Zjet$ and diboson processes, and are estimated
with MC samples. The final estimates for all backgrounds are given in Table~\ref{tab:Yields},
along with the expected signal.

\begin{table}[tb]
\begin{center}
{ 
\begin{tabular}{l|!!|!} 
\hline \hline
  & \multicolumn{2}{c|}{Resolved channel} & \multicolumn{1}{c}{Merged channel} \\
  & \multicolumn{1}{c}{$e^+$ and $\mu^+$} &  \multicolumn{1}{c|}{$e^-$ and $\mu^-$}  & \multicolumn{1}{c}{$e$ and $\mu$} \\
\hline
 $\Wjet$ & 92,37  & 51,29 & 19.4,9.9   \\
 \ttbar & 59,18  & 63,35 & 6.8,2.8   \\
 Single-top & 10.0,5.6  &  5.5,3.2 & 2.2,1.2   \\
 Diboson & 8.6,5.7  & 10.8,6.4 & 1.6,1.2  \\
 $\Zjet$ & 4.5,1.5  & 3.4,2.4 & 0.58,0.64  \\
 Multijet & 16,16  & 12,12 & 1.8,1.9  \\
\hline
Total background & 190,53 & 145,54 & 32,12 \\
\hline
\signal\ (SM)  & 3.66,0.82 &  2.34,0.56 & 0.54,0.22   \\
\signal\ ($\alpha_4=0.1, \alpha_5=0$)   & 21.0,4.2  & 9.2,1.9 &  15.1,4.4 \\
\hline\hline
Data & \multicolumn{1}{c}{173}  & \multicolumn{1}{c}{131}  & \multicolumn{1}{c}{32} \\
\hline \hline
\end{tabular}
}
\caption{The expected number of events passing the final event
selection, together with the number of events observed in data.  
The expected \signal\ contributions for a representative point in the 
aQGC parameter space  ($\alpha_4=0.1, \alpha_5=0$) and for
the SM ($\alpha_4=\alpha_5=0$) are shown for comparison.
The quoted errors include
all systematic uncertainties in the expected yields.  The error in the total background
is computed including correlations between the various background components.
}
\label{tab:Yields}
\end{center}
\end{table}

\begin{figure}[htbp]
\begin{center}
\subfigure[]{
 \includegraphics[width=0.47\textwidth,keepaspectratio]{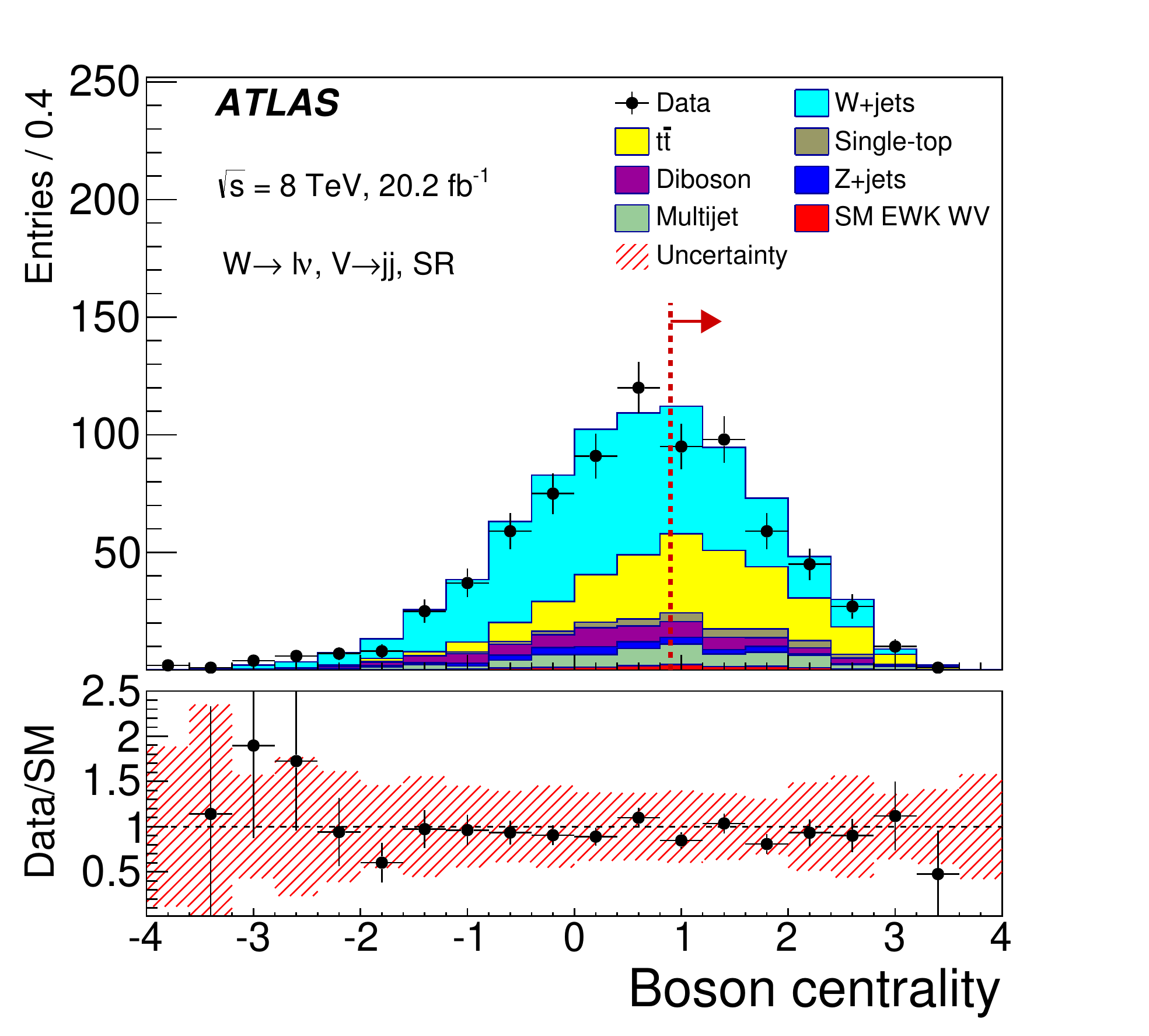}
}
\subfigure[]{
 \includegraphics[width=0.47\textwidth,keepaspectratio]{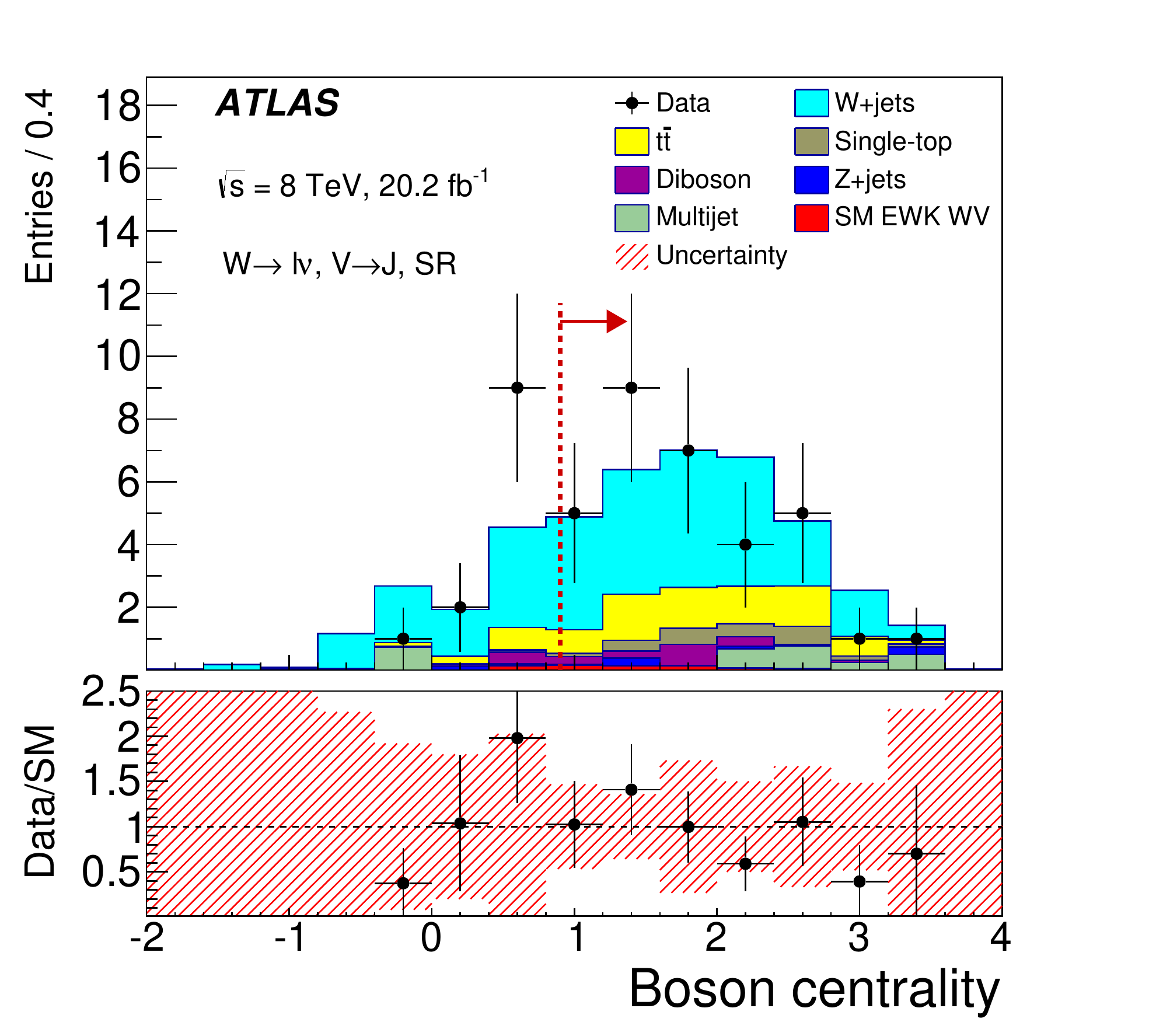}
}
\subfigure[]{
 \includegraphics[width=0.47\textwidth,keepaspectratio]{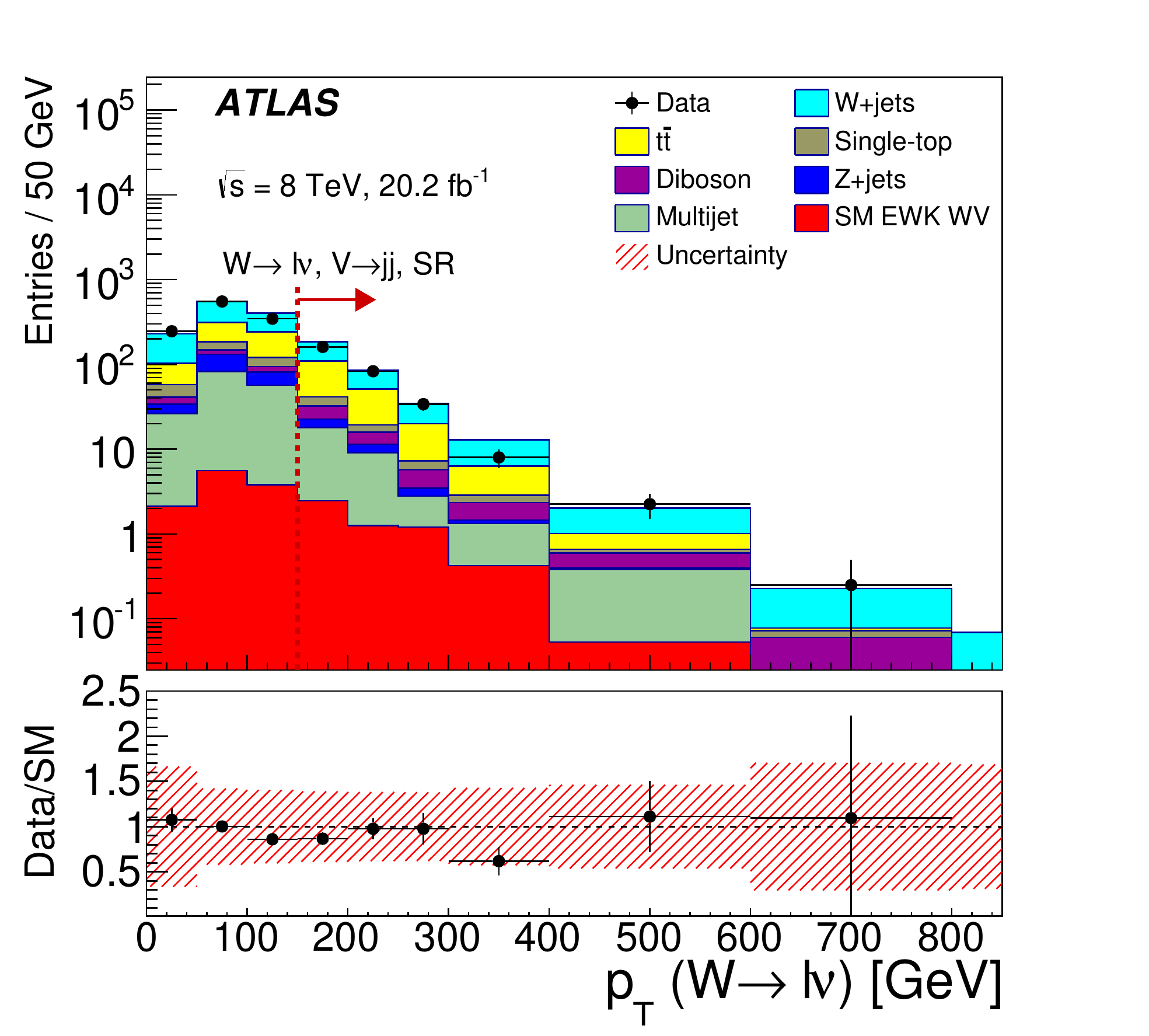}
}
\subfigure[]{
 \includegraphics[width=0.47\textwidth,keepaspectratio]{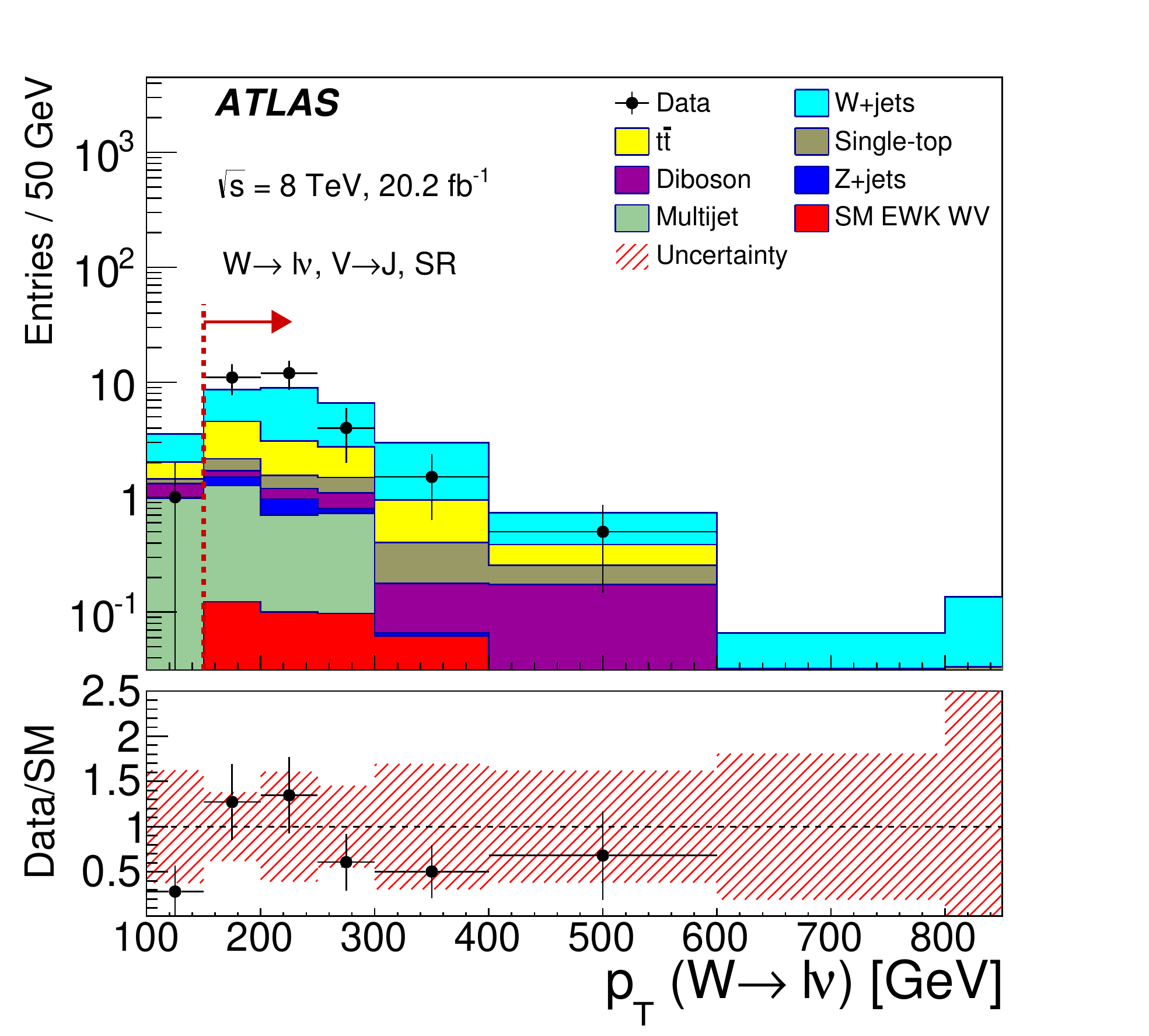}
}
\caption{
The observed boson centrality (top) and $\pTWlep$ (bottom) distributions, compared to the
SM prediction.  Plots (a) and (c) show the \resolved\ ($\Vtojj$) signal region (SR), and plots (b) and (d) show
the \merged\ ($\VtoJ$) signal region, except that the $\bosCen>\cutcentralR$ requirement is not applied
for the boson centrality plots, and the $\pTWlep>150~\gev$ requirement is not applied for the $\pTWlep$
plots.  The red vertical lines and arrows indicate the signal region selection.
The last bin includes overflow.
}
\label{fig:n-1PaperPlots}
\end{center}
\end{figure}

The background modeling is further cross-checked in Figure~\ref{fig:n-1PaperPlots}, which shows 
data--MC comparisons of the $\pTWlep$ and boson centrality distributions.  In these plots,
all of the signal-region selection criteria are applied, except for the selection criterion
for the variable ($\pTWlep$ or boson centrality) being plotted. The data agree with 
the predictions within the systematic uncertainty bands.

%-------------------------------------------------------------------------------

%-------------------------------------------------------------------------------
\section{Systematic uncertainties}
\label{sec:systematics}
A variety of sources of systematic uncertainty are 
considered.  The effect of systematic uncertainties in the 
background and signal rates, and in the shape of the $\mTWV$ 
distribution of background and signal events, are accounted
for. 

Systematic uncertainties in the jet energy scale (JES) and jet
energy resolution (JER) are calculated separately 
for small-$R$~\cite{PERF-2011-03,PERF-2011-04} and large-$R$
jets.  For the large-$R$ jets, uncertainties in the
jet mass scale and jet mass resolution are included and 
account for uncertainty in the modeling of the jet substructure.
The large-$R$ jet energy and mass scale uncertainties are derived from
ratios of calorimeter-jets to track-jets and from 
$\gamma$+jet balance studies.  The large-$R$ jet energy and mass resolution
uncertainties are estimated by applying a smearing factor so that the 
resolutions increase by a factor of $20\%$; this uncertainty is based on 
previous studies of large-$R$ jets~\cite{STDM-2011-19,HIGG-2013-19}.
The jet-related uncertainties are the most significant detector-related
uncertainties in the analysis.

Uncertainties in lepton reconstruction and identification,
soft terms entering the $\met$ calculation, and $b$-tagging
are accounted for, and have a minor effect.  The uncertainty 
in the integrated luminosity is also included~\cite{Aaboud:2016hhf}.

Systematic uncertainties in the signal model are taken into
account, including variations in the model of fragmentation,
parton shower, and hadronization; factorization and 
renormalization scales; and the PDFs.
Uncertainties in the $\Vjet$ background model are accounted
for by varying the factorization and renormalization scales,
and the scale for matching matrix elements 
to parton showers~\cite{Hoeche:2009rj}. 
The full difference between the
data-driven $\Wjet$ scale factor and 1.00 is also included
as an uncertainty: $0.82\pm0.18$; this scale factor is varied independently in
each of the three event categories.
Uncertainties in the $\ttbar$ modeling are estimated by varying
the matrix-element generator, the fragmentation/parton-shower/hadronization 
model, and the amount of initial-state and final-state radiation.
A 100\% uncertainty is applied to the multijet background prediction,
and covers uncertainties in the data-driven estimation procedure.
For the single-top-quark, diboson, and electroweak $\Wjet$ predictions,
instead of computing separate modeling uncertainties from
individual sources,
an overall normalization uncertainty of 50\% is applied, which is 
taken as an 
estimate of their modeling uncertainties based on studies of 
other background processes. The uncertainties in the
multijet, single-top-quark, diboson, and electroweak $\Wjet$ backgrounds
only increase the overall background uncertainty by about $2-3\%$.

There is also a statistical uncertainty in the expected number of background and signal
in each bin of $\mTWV$, due to the size of the MC samples and the 
numbers of events 
in the multijet control regions.

The uncertainties in the total background are dominated by jet uncertainties
and $\Vjet$ modeling, and are summarized in Table~\ref{tab:systNormSummary}.
The uncertainty in the signal yield is about 20\% (30\%) in the resolved (merged) categories, 
and is dominated by the signal model variations and the jet uncertainties.

\begin{table}[tb]
\begin{center}
{ 
\begin{tabular}{l|c|c} 
\hline \hline
   & \multicolumn{2}{c}{Fractional uncertainty} \\
\hline
Source & {\em Resolved} & {\em Merged}  \\
\hline
$\Vjet$ modeling & 0.13 &  0.29  \\  
$\ttbar$ modeling & 0.14 & 0.07  \\
Multijet yield & 0.06 & 0.05  \\
Minor background yields & 0.04 &  0.05  \\
Jet reconstruction & 0.21 &  0.17  \\
Other detector/luminosity & 0.04 &  0.03  \\
Limited stats in MC or CR & 0.02  & 0.06 \\
\hline
Total & 0.29  &  0.36  \\
\hline \hline
\end{tabular}
}
\caption{ Summary of the fractional uncertainty in the total background yields in the signal region, broken down into 
different categories of systematic uncertainties. 
}
\label{tab:systNormSummary}
\end{center}
\end{table}

%-------------------------------------------------------------------------------

%-------------------------------------------------------------------------------
\section{Results}
\label{sec:result}
%-------------------------------------------------------------------------------

\begin{figure}[htbp]
\begin{center}
\subfigure[]{
 \includegraphics[width=0.47\textwidth,keepaspectratio]{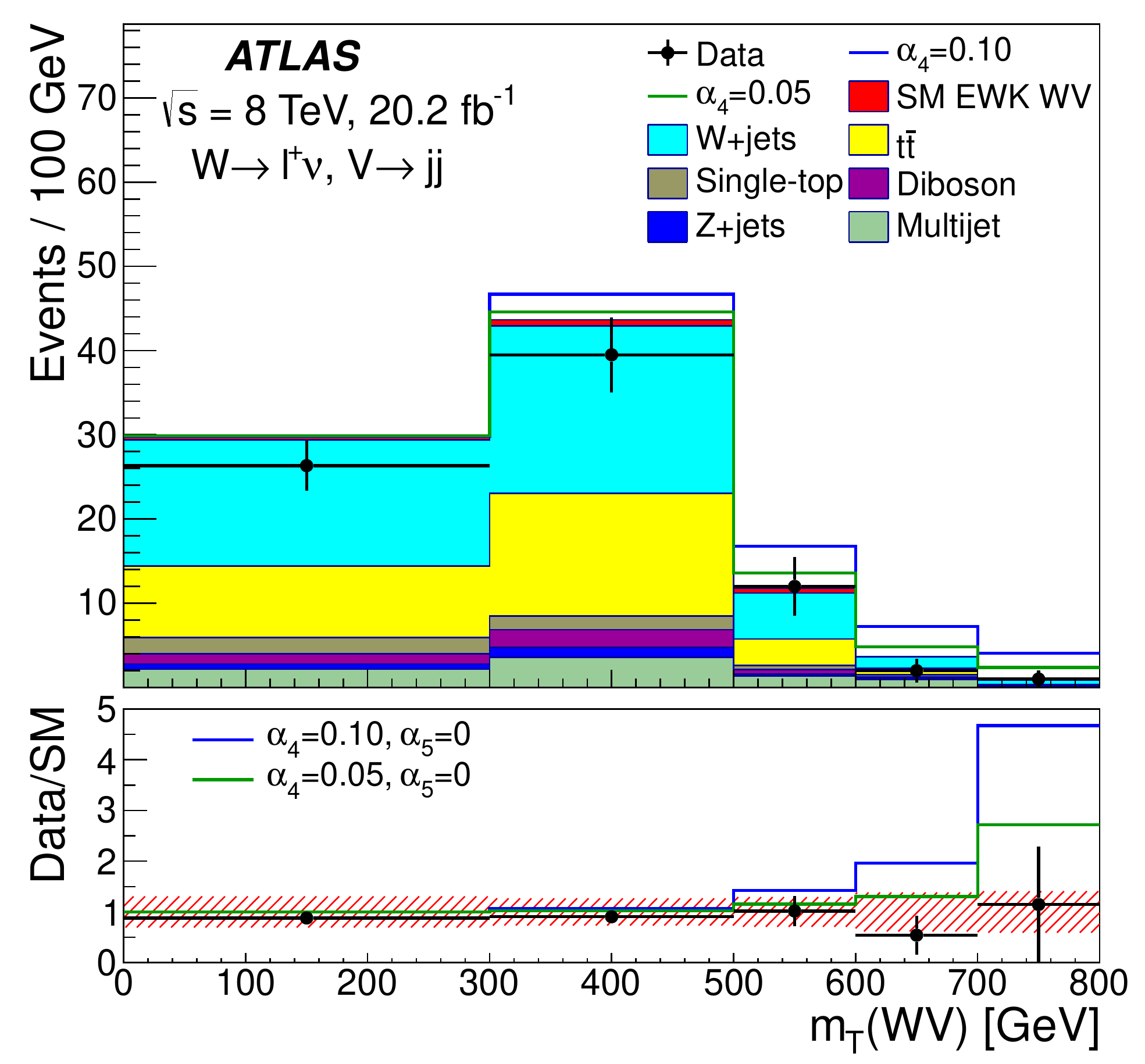}
}
\subfigure[]{
 \includegraphics[width=0.47\textwidth,keepaspectratio]{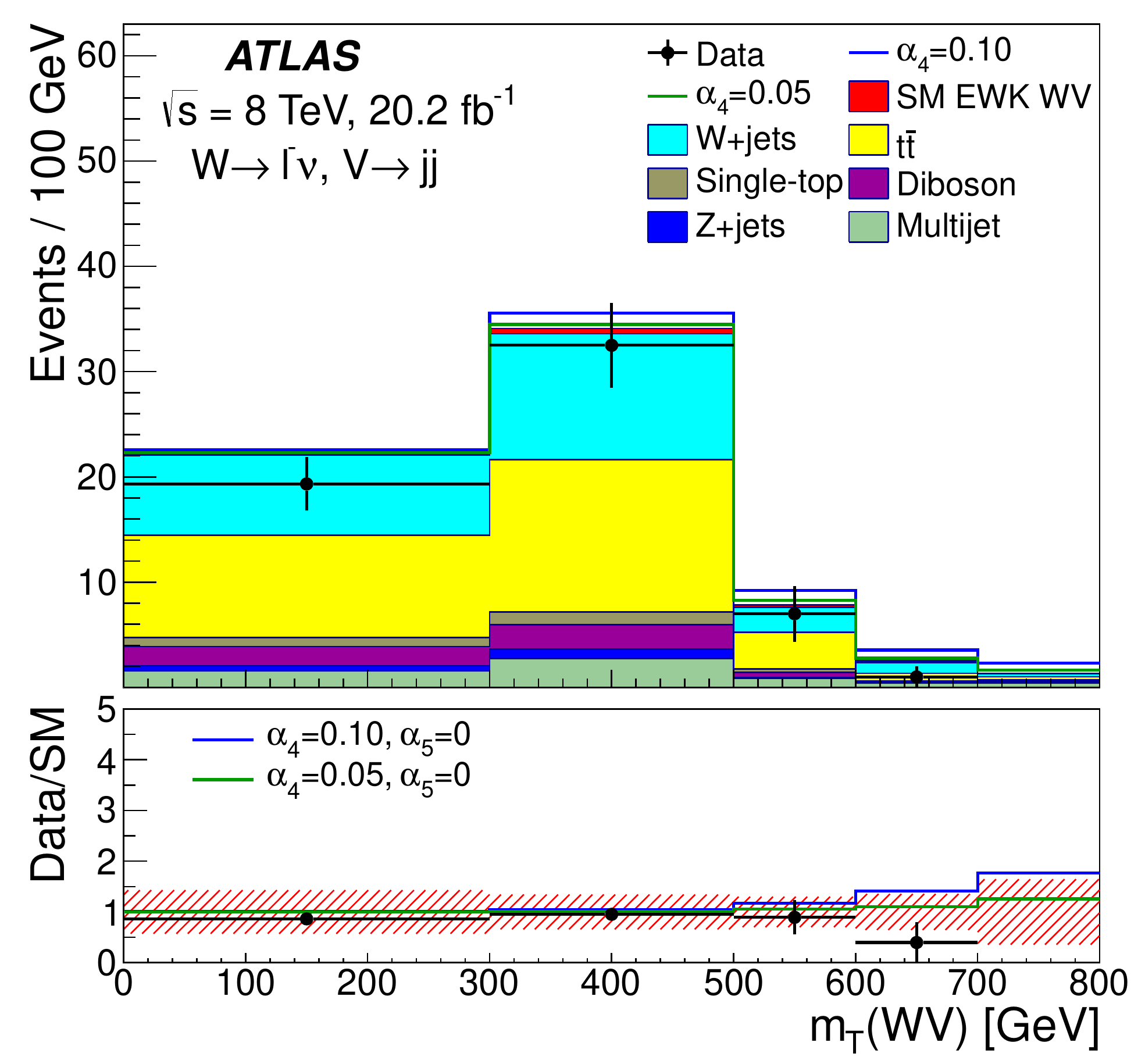}
}
\subfigure[]{
 \includegraphics[width=0.47\textwidth,keepaspectratio]{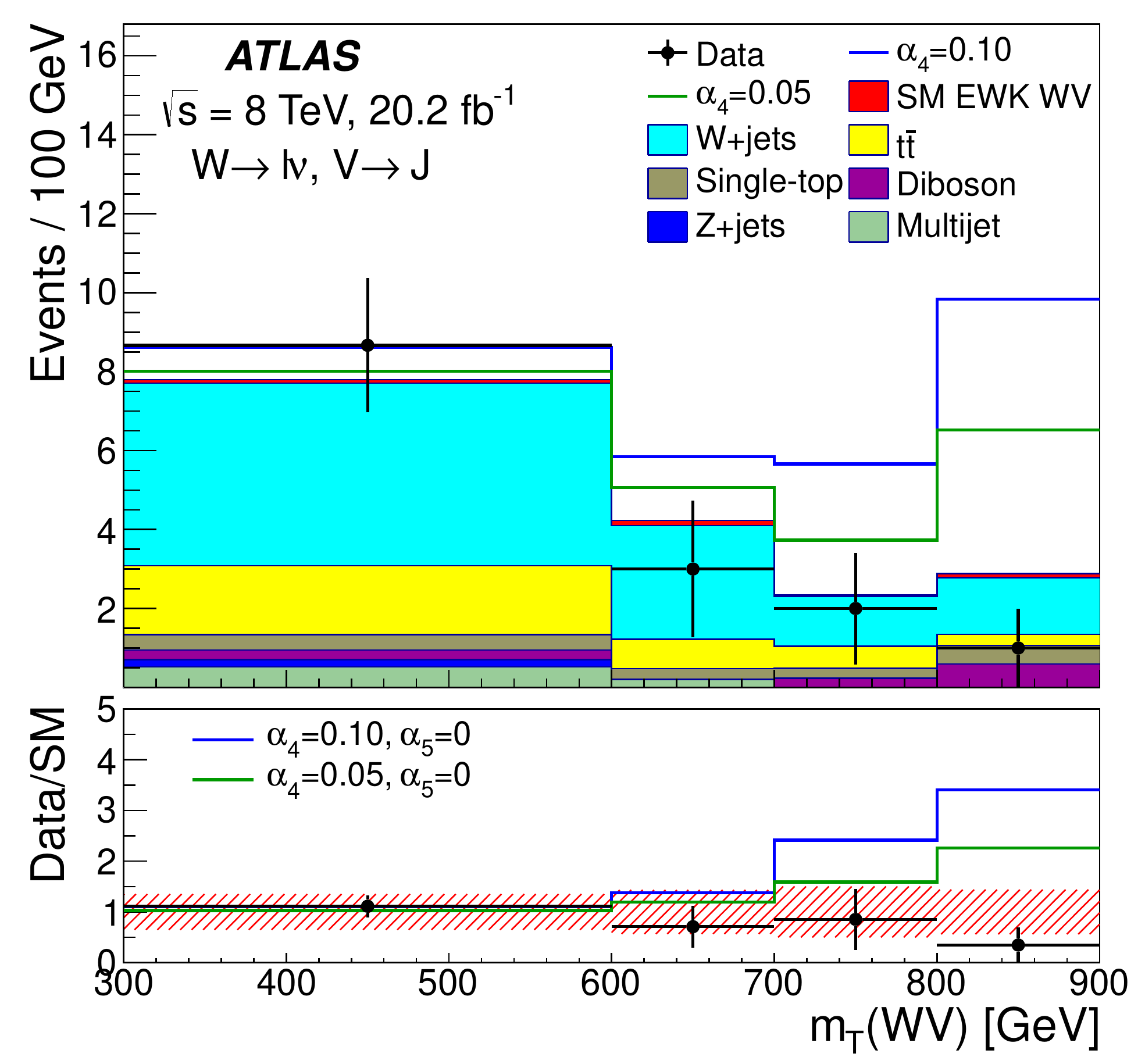}
}
\caption{
The observed $\mTWV$ distribution, overlaid with background and \signal\ prediction, 
after applying the full selection. The expected 
enhancements due to aQGC values of ($\alpha_4=0.1$, $\alpha_5=0$) 
and ($\alpha_4=0.05$, $\alpha_5=0$) are also shown.  The plotted regions are 
(a) the \resolved\ ($\Vtojj$) region, $e^+$ and $\mu^+$ combined;  
(b) the \resolved\ region, $e^-$ and $\mu^-$ combined; and
(c) the \merged\ ($\VtoJ$) region, $e^+$, $e^-$, $\mu^+$, and $\mu^-$ combined.
The last bin includes overflow.
}
\label{fig:aQGCfitDistObserved}
\end{center}
\end{figure}

A search for aQGC contributions is performed by examining the $\mTWV$
distribution of events that satisfy the full selection.  
The $\mTWV$ distribution of events is shown 
in Figure~\ref{fig:aQGCfitDistObserved}, split up into the 
three categories defined in Section~\ref{sec:eventselection}.
 The enhancements of \signal\ expected for different aQGC
values are shown for comparison.  No evidence of an aQGC is
observed in the data, so the allowed 95\% confidence intervals 
are computed for the 
aQGC parameters $\alpha_4$ and $\alpha_5$.  

\begin{figure}[htbp]
\begin{center}
 \includegraphics[width=0.47\textwidth,keepaspectratio]{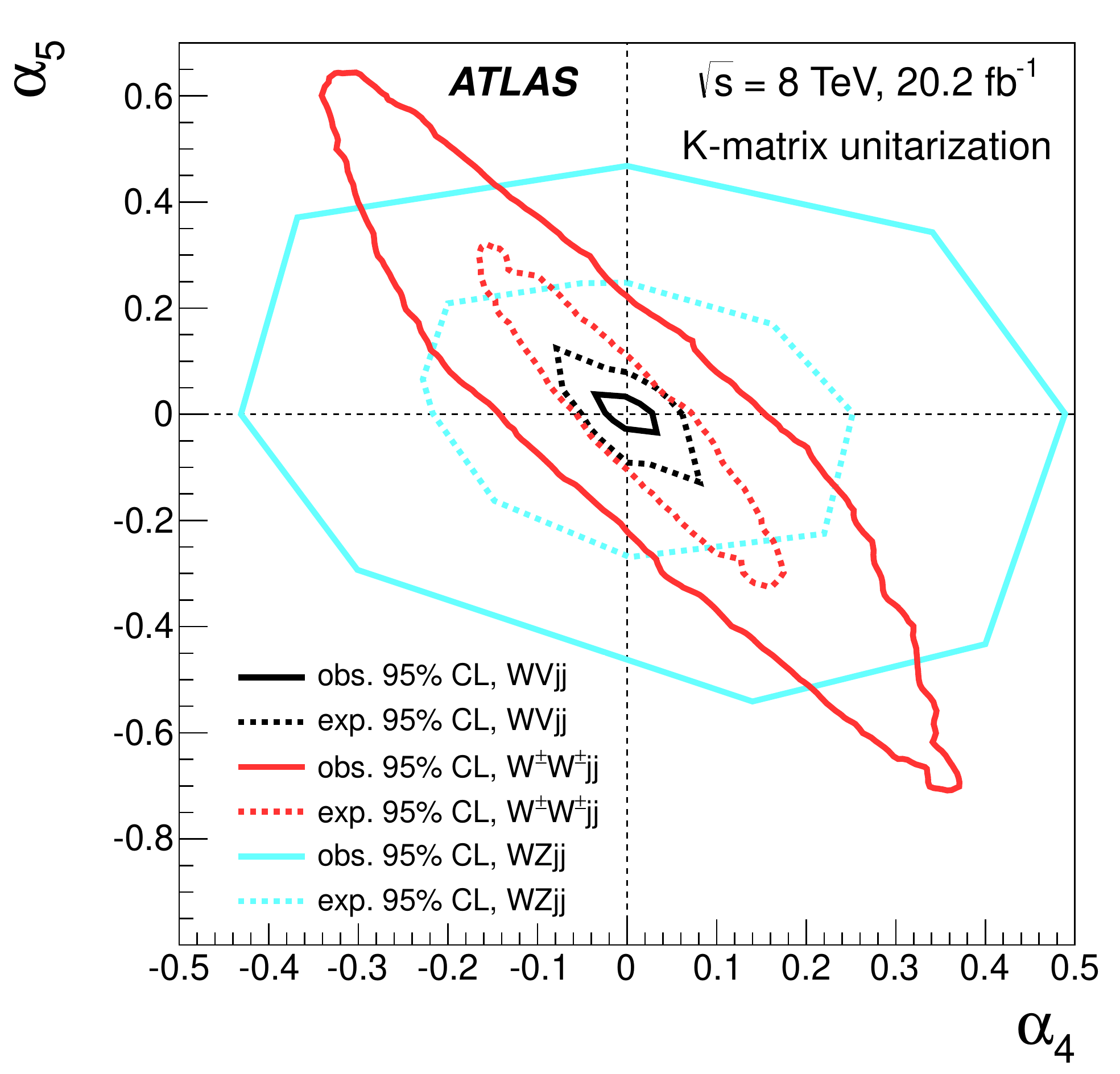}
\caption{
The observed 2D confidence region (solid black contour) for $\alpha_4$ and $\alpha_5$, at 95\% CL.  
The expected 2D confidence region (dotted black contour) is also shown, computed using the Asimov dataset~\cite{Cowan:2010js}.   
Results from this analysis (in black) are compared
to observed and expected confidence regions from previous ATLAS analyses of 
$\ssWW$~\cite{ssWW8TeVPRD} (in red) and $WZ$~\cite{Aad:2016ett} (in cyan) VBS production.
}
\label{fig:2dlimits}
\end{center}
\end{figure}

The confidence intervals on  $\alpha_4$ and $\alpha_5$ are calculated by using a binned
profile-likelihood~\cite{Cowan:2010js} fit to the $\mTWV$ distribution in the three event categories.
  Systematic uncertainties are incorporated into the fit using 28 nuisance parameters.
The frequentist 95\% confidence level (CL) intervals are computed using pseudo-experiments.  For each 
aQGC point, the ratio of the likelihood to the likelihood of 
the best-fit aQGC point is calculated.  An 
aQGC point is excluded at 95\% CL if at least 95\% of the random pseudo-experiments
have a profile-likelihood ratio greater than the observed one.
At 95\% CL, the observed confidence intervals are $\aFourLoObs<\alpha_4<\aFourHiObs$
and $\aFiveLoObs<\alpha_5<\aFiveHiObs$, where the confidence interval on each parameter is
calculated while fixing the other parameter to zero.  The expected 95\% confidence intervals
are $\aFourLoExp<\alpha_4<\aFourHiExp$ and $\aFiveLoExp<\alpha_5<\aFiveHiExp$.
The observed confidence intervals are stronger than expected; under the SM hypothesis, there is
a $12-15\%$ probability of obtaining confidence intervals  more stringent than the observed ones.
The expected and observed confidence intervals are summarized in Table~\ref{tab:1dintervals}. 
This table also shows the $1-$ and $2-$sigma uncertainty bands on the expected confidence intervals.  These
uncertainty bands show that the measured confidence intervals can vary significantly from pseudo-experiment
to pseudo-experiment; this behavior is expected since most of the sensitivity to the aQGC parameters comes from 
high-$\mTWV$ bins with few events and large uncertainties.
The two-dimensional (2D) confidence region for $\alpha_4$ and $\alpha_5$ is shown in Figure~\ref{fig:2dlimits}.  
The observed $\alpha_4$ and $\alpha_5$ confidence intervals are more stringent than
existing confidence intervals for these parameters, which are obtained from VBS 
$\ssWWlep$~\cite{ssWW8TeVPRD} and $\WZlep$~\cite{Aad:2016ett} measurements from ATLAS.

\begin{table}[tb]
\begin{center}
{ 
\begin{tabular}{l|c|c|c|c} 
\hline \hline
   & Expected & Expected $\pm1\sigma$ & Expected  $\pm2\sigma$ & Observed \\
\hline
lower limit, $\alpha_4$ & $\aFourLoExp$ & $[\aFourLoExpMinusOneSig, \aFourLoExpPlusOneSig]$  & $[\aFourLoExpMinusTwoSig, \aFourLoExpPlusTwoSig]$ & $\aFourLoObs$  \\
upper limit, $\alpha_4$ & $\aFourHiExp$ & $[\aFourHiExpMinusOneSig, \aFourHiExpPlusOneSig]$  & $[\aFourHiExpMinusTwoSig, \aFourHiExpPlusTwoSig]$ & $\aFourHiObs$  \\
lower limit, $\alpha_5$ & $\aFiveLoExp$ & $[\aFiveLoExpMinusOneSig, \aFiveLoExpPlusOneSig]$  & $[\aFiveLoExpMinusTwoSig, \aFiveLoExpPlusTwoSig]$ & $\aFiveLoObs$  \\
upper limit, $\alpha_5$ & $\aFiveHiExp$ & $[\aFiveHiExpMinusOneSig, \aFiveHiExpPlusOneSig]$  & $[\aFiveHiExpMinusTwoSig, \aFiveHiExpPlusTwoSig]$ & $\aFiveHiObs$  \\
\hline \hline
\end{tabular}
}
\caption{The observed and expected lower and upper limits of the 95\% confidence intervals for $\alpha_4$ and $\alpha_5$.  The $\pm1\sigma$ and $\pm2\sigma$ 
uncertainty bands on the expected lower and upper limits are also shown for comparison.  The $\alpha_4$ confidence intervals are computed while fixing $\alpha_5$
to zero, and vice-versa.
}
\label{tab:1dintervals}
\end{center}
\end{table}

 The use of the 
``\merged'' category of events significantly improves the aQGC sensitivity of the analysis because
most of the aQGC sensitivity comes from the highest-$\mTWV$ bins, where the \merged\ category is powerful.
The expected aQGC confidence intervals are about 40\% more stringent when including this category than when only using the
\resolved\ events.

% \FloatBarrier

%-------------------------------------------------------------------------------
\section{Conclusions}
\label{sec:conclusion}
%-------------------------------------------------------------------------------
A search is performed for anomalous quartic gauge couplings in
$WW$ and $WZ$ production via vector-boson scattering.  The analysis is performed
with $\ourLumiNoError$ of ATLAS data from  $\rts=8~\tev$   $pp$ collisions at the LHC.

The search is based on a signature of $W(\ell\nu)\Vqq$ plus two jets with a high
dijet invariant mass. The $\Vqq$ system is reconstructed either as two separate jets
 or as a single, large-radius jet, making use of jet substructure techniques. 
A search phase space is used that is designed to be particularly sensitive to aQGCs, 
and is based on event topology, the $V$ decay angle, and high transverse momentum.

No excess is seen in the data, and so limits are placed on aQGC parameters by fitting
the diboson transverse-mass distribution.  At 95\% CL, the observed limits are
$\aFourLoObs<\alpha_4<\aFourHiObs$ and $\aFiveLoObs<\alpha_5<\aFiveHiObs$.
These limits are more stringent than the previous constraints on these 
parameters, obtained in searches for vector-boson scattering in the 
$\ssWWlep$ and $\WZlep$ channels.  
This result demonstrates that a semileptonic
channel can have strong experimental sensitivity to new physics contributions to
vector-boson scattering.

%-------------------------------------------------------------------------------
\section*{Acknowledgements}
%-------------------------------------------------------------------------------
% Acknowledgements for papers with collision data
% Version 2-Aug-2016

% Standard acknowledgements start here
%----------------------------------------------
We thank CERN for the very successful operation of the LHC, as well as the
support staff from our institutions without whom ATLAS could not be
operated efficiently.

We acknowledge the support of ANPCyT, Argentina; YerPhI, Armenia; ARC, Australia; BMWFW and FWF, Austria; ANAS, Azerbaijan; SSTC, Belarus; CNPq and FAPESP, Brazil; NSERC, NRC and CFI, Canada; CERN; CONICYT, Chile; CAS, MOST and NSFC, China; COLCIENCIAS, Colombia; MSMT CR, MPO CR and VSC CR, Czech Republic; DNRF and DNSRC, Denmark; IN2P3-CNRS, CEA-DSM/IRFU, France; GNSF, Georgia; BMBF, HGF, and MPG, Germany; GSRT, Greece; RGC, Hong Kong SAR, China; ISF, I-CORE and Benoziyo Center, Israel; INFN, Italy; MEXT and JSPS, Japan; CNRST, Morocco; FOM and NWO, Netherlands; RCN, Norway; MNiSW and NCN, Poland; FCT, Portugal; MNE/IFA, Romania; MES of Russia and NRC KI, Russian Federation; JINR; MESTD, Serbia; MSSR, Slovakia; ARRS and MIZ\v{S}, Slovenia; DST/NRF, South Africa; MINECO, Spain; SRC and Wallenberg Foundation, Sweden; SERI, SNSF and Cantons of Bern and Geneva, Switzerland; MOST, Taiwan; TAEK, Turkey; STFC, United Kingdom; DOE and NSF, United States of America. In addition, individual groups and members have received support from BCKDF, the Canada Council, CANARIE, CRC, Compute Canada, FQRNT, and the Ontario Innovation Trust, Canada; EPLANET, ERC, FP7, Horizon 2020 and Marie Sk{\l}odowska-Curie Actions, European Union; Investissements d'Avenir Labex and Idex, ANR, R{\'e}gion Auvergne and Fondation Partager le Savoir, France; DFG and AvH Foundation, Germany; Herakleitos, Thales and Aristeia programmes co-financed by EU-ESF and the Greek NSRF; BSF, GIF and Minerva, Israel; BRF, Norway; Generalitat de Catalunya, Generalitat Valenciana, Spain; the Royal Society and Leverhulme Trust, United Kingdom.

The crucial computing support from all WLCG partners is acknowledged gratefully, in particular from CERN, the ATLAS Tier-1 facilities at TRIUMF (Canada), NDGF (Denmark, Norway, Sweden), CC-IN2P3 (France), KIT/GridKA (Germany), INFN-CNAF (Italy), NL-T1 (Netherlands), PIC (Spain), ASGC (Taiwan), RAL (UK) and BNL (USA), the Tier-2 facilities worldwide and large non-WLCG resource providers. Major contributors of computing resources are listed in Ref.~\cite{ATL-GEN-PUB-2016-002}.
%----------------------------------------------
%-------------------------------------------------------------------------------

\printbibliography

\newpage
% ATLAS Collaboration author list
% Data extracted on 22-Nov-2016 for paper reference STDM-2015-09
%\documentclass[11pt]{article}
%\usepackage{a4wide}\begin{document}
\begin{flushleft}
{\Large The ATLAS Collaboration}

\bigskip

M.~Aaboud$^\textrm{\scriptsize 136d}$,
G.~Aad$^\textrm{\scriptsize 87}$,
B.~Abbott$^\textrm{\scriptsize 114}$,
J.~Abdallah$^\textrm{\scriptsize 8}$,
O.~Abdinov$^\textrm{\scriptsize 12}$,
B.~Abeloos$^\textrm{\scriptsize 118}$,
R.~Aben$^\textrm{\scriptsize 108}$,
O.S.~AbouZeid$^\textrm{\scriptsize 138}$,
N.L.~Abraham$^\textrm{\scriptsize 152}$,
H.~Abramowicz$^\textrm{\scriptsize 156}$
% ...
\bigskip
\\
% ...
$^{8}$ Department of Physics, The University of Texas at Arlington, Arlington TX, United States of America\\
% ...
$^{12}$ Institute of Physics, Azerbaijan Academy of Sciences, Baku, Azerbaijan\\
% ...
$^{87}$ CPPM, Aix-Marseille Universit{\'e} and CNRS/IN2P3, Marseille, France\\
% ...
$^{108}$ Nikhef National Institute for Subatomic Physics and University of Amsterdam, Amsterdam, Netherlands\\
% ...
$^{114}$ Homer L. Dodge Department of Physics and Astronomy, University of Oklahoma, Norman OK, United States of America\\
% ...
$^{118}$ LAL, Univ. Paris-Sud, CNRS/IN2P3, Universit{\'e} Paris-Saclay, Orsay, France\\
% ...
$^{136}$ $^{(a)}$ Facult{\'e} des Sciences Ain Chock, R{\'e}seau Universitaire de Physique des Hautes Energies - Universit{\'e} Hassan II, Casablanca; $^{(b)}$ Centre National de l'Energie des Sciences Techniques Nucleaires, Rabat; $^{(c)}$ Facult{\'e} des Sciences Semlalia, Universit{\'e} Cadi Ayyad, LPHEA-Marrakech; $^{(d)}$ Facult{\'e} des Sciences, Universit{\'e} Mohamed Premier and LPTPM, Oujda; $^{(e)}$ Facult{\'e} des sciences, Universit{\'e} Mohammed V, Rabat, Morocco\\
% ...
$^{138}$ Santa Cruz Institute for Particle Physics, University of California Santa Cruz, Santa Cruz CA, United States of America\\
% ...
$^{152}$ Department of Physics and Astronomy, University of Sussex, Brighton, United Kingdom\\
% ...
$^{156}$ Raymond and Beverly Sackler School of Physics and Astronomy, Tel Aviv University, Tel Aviv, Israel\\
% ...
$^{a}$ Also at Department of Physics, King's College London, London, United Kingdom\\
$^{b}$ Also at Institute of Physics, Azerbaijan Academy of Sciences, Baku, Azerbaijan\\
$^{c}$ Also at Novosibirsk State University, Novosibirsk, Russia\\
$^{d}$ Also at TRIUMF, Vancouver BC, Canada\\
$^{e}$ Also at Department of Physics {\&} Astronomy, University of Louisville, Louisville, KY, United States of America\\
% ...
\end{flushleft}

% Created with xml2latex.py

%-------------------------------------------------------------------------------
\clearpage
%\appendix
%\part*{Auxiliary material}
%\FloatBarrier
%\addcontentsline{toc}{part}{Auxiliary material}
%---------------------------------------------------
%\input{Auxiliary}

\end{document}